\documentclass{sigplanconf}

\usepackage{amsmath}
\usepackage{url}
\usepackage{graphicx}
\begin{document}
\conferenceinfo{CloudDP '13,} {April 14, 2013, Prague, Czech Republic.}
\CopyrightYear{2013}
\crdata{978-1-4503-2075-7}
\title{Experiences of Using a Hybrid Cloud to Construct an Environmental Virtual Observatory}
\authorinfo{Yehia Elkhatib \and Gordon S. Blair}
           {School of Computing \& Communications\\
	Lancaster University, LA1 4YD, United Kingdom}
           {\{y.elkhatib $|$ g.blair\}@lancaster.ac.uk}

\authorinfo{Bholanathsingh Surajbali}
           {CAS Software AG\\
	CAS-Weg 1-5, Karlsruhe, Germany}
           {b.surajbali@cas.de}

\maketitle

\begin{abstract}
Environmental science is often fragmented: data is collected using mismatched formats and conventions, and models are misaligned and run in isolation. Cloud computing offers a lot of potential in the way of resolving such issues by supporting data from different sources and at various scales, by facilitating the integration of models to create more sophisticated software services, and by providing a sustainable source of suitable computational and storage resources. In this paper, we highlight some of our experiences in building the Environmental Virtual Observatory pilot (EVOp), a tailored cloud-based infrastructure and associated web-based tools designed to enable users from different backgrounds to access data concerning different environmental issues. We review our architecture design, the current deployment and prototypes. We also reflect on lessons learned. We believe that such experiences are of benefit to other scientific communities looking to assemble virtual observatories or similar virtual research environments.
\end{abstract}




\section{Introduction}

Environmental science is studied by different communities in the academic, governmental and commerical sectors for different purposes. Each community forms its own data management conventions, such as decisions about spatial and temporal resolution, storage format, dataset description, method of access, etc. Many of these communities work in isolation, hence it is not unexpected that their conventions vary significantly from each other. Such misalignment is a major hindrance to data discovery and use, complicating any sort of collaboration between different communities on a common environmental topic.

One way to transcend such gaps is to introduce a shared space to facilitate collaboration. We define such space as a \textit{virtual observatory}, where different stakeholders would be able to view and add environmental resources such as observation datasets, analysis results, annotation and analysis processes, and visualisation tools. For such a virtual observatory to be realised, an infrastructure is needed to support data from different sources and at various scales, to integrate processes in order to create more sophisticated and collaborative software services, and to provide a sustainable source of adequate computational and storage resources.

Cloud computing offers great potential for such capacities. Cloud resources are easy to forge and steer to better serve the needs of an application. Such flexibility presents capabilities to integrate varied resources (i.e. processes and data) used by different communities. It also helps transcend traditional software stacks, circumventing a lot of the development restrictions and deployment difficulties imposed by previous distributed systems.

The Environmental Virtual Observatory pilot (EVOp) project \cite{blair2011cloud} is funded by the UK Natural Environment Research Council to support the assembly of advanced cloud-based services that can benefit a number of communities with different interests in pressing environmental issues. For the pilot phase (2011-2012), the project focused on water related environmental disciplines including hydrology, land cover management, and diffuse pollution. 

EVOp is designed for use without any programming prerequisites by both domain specialists and non-specialists. For environmental scientists, it allows them to worry less about some of the repetitive non-scientific tasks that they have to do as part of their work. For policy makers, EVOp serves as a decision support system: it provides open access to explore current data and tools to investigate the effect of new policies. For members of local communities, EVOp enables them to explore the impact of different practices relating to farming, water management, etc. For the general public, EVOp supports raising general awareness of environmental issues, encouraging a wider discussion about impact in addition to individual and collective contribution.

This paper offers some experiences in building the EVOp infrastructure, and in migrating data and tools to the cloud. We believe that the architectural principles, migration processes, and operational measures of EVOp are shared with other scientific communities looking to assemble similar cloud-based virtual research environments. Hence, many of our learned lessons and faced difficulties would be of benefit to a much wider audience.

The rest of the paper is structured as follows. Section \ref{sec-arch} introduces the design of the EVOp infrastructure, then Section \ref{sec-deploy} presents its current deployment. Section \ref{sec-reflections} reflects on the main difficulties faced and lessons learned. Section \ref{sec-future} discusses future steps of EVOp, while Section \ref{sec-related} highlights some related work. Finally, Section \ref{sec-conclusion} offers concluding remarks.

\section{Architecture}
\label{sec-arch}

\subsection{Design Concepts}
\label{sec-arch-concepts}
The design of the EVOp infrastructure adopts the principles of cloud computing and associated technologies to deliver a system of low operational costs at the infrastructure level and high flexibility at the application level. The main concepts behind the design of the EVOp infrastructure are now discussed.

\subsubsection{Everything as a Service (XaaS)}
\label{sec-arch-concepts-XaaS}
All system resources (such as datasets and analysis processes) are accessible via web service interfaces. This hides from the users details of where resources are held and how they are managed. Such abstraction translates to a better user experience as complicated issues are offloaded allowing the users to focus on solving domain-specific problems. XaaS also means that models and simulations could process datasets without necessarily giving them away, avoiding some of the thorny issues of data ownership. At the hardware level, XaaS abstracts away the complexities of distribution, reliability and availability.
  
\subsubsection{RESTful Web Services}
\label{sec-arch-concepts-REST}
In addition to the previous point, all web services interfaces are of a uniform view, designed according to the Representational State Transfer (REST) architectural principles \cite{fielding2000architectural}. In contrast to SOAP or \textit{big} web services, the REST architectural style is resource-oriented rather than transaction-oriented. Hence, RESTful web services remain completely stateless with all data required to transition between different states being included in the service request. 

\subsubsection{Virtualization}
\label{sec-arch-concepts-virt}
Virtualization brings dynamic provisioning of bespoke environments where everything from the hardware, platform, libraries, etc. can be customised to suit the exact needs of an application. Such flexibility is of great value as it enables scientists to move their computations to the cloud with very few restrictions. 

\subsubsection{Federated Cloud}
\label{sec-arch-concepts-federated}
We use a hybrid infrastructure comprised of both private and public cloud resources in order to rise above the shortcomings of one solution and to control upfront and running expenditures. In an effort to promote portability and to avoid being tied in to one provider, we decided to use the cross-cloud library jclouds \cite{jclouds}.

\subsubsection{2-Tier not 3-Tier}
\label{sec-arch-concepts-tiers}
The three-tier design model is a widely adopted one for obvious reasons: it provides separation of concerns between the levels of data management, processing or business logic, and presentation. Tiers could be implemented, managed, and scaled independently which is an important consideration for the deployment of large scale systems. The trade-off is the need for measures to avoid performance bottlenecks between levels (e.g. caching) \cite{Malkowski09workload} and to maintain consistency across each level. In a sense, functional decomposition introduced by the three-tiered architecture alleviates some software development complexities but replaces them with operations and system maintenance difficulties.

For the purposes of EVOp, we identified that the majority of processes need access only to historical data that is read a lot but rarely written (once initially and then appended to every once in a while). Such data does not need to be made available through a shared system. Instead, it could easily be provided to running cloud instances as ephemeral disk storage. This allows instances to have all required data at a predictable performance level that is independent of the overall system demand for that data. Therefore, we decide to couple data and modelling logic where possible.

\subsection{Design}
\label{sec-arch-design}
The EVOp portal contains a mixture of data sources that a user can explore. These include live data (such as live river level, temperature, etc.), historical data (e.g. rainfall) and others (e.g. webcam images). Some are managed by the EVOp team while other information resources are mashups of external data sources.

\begin{figure}[!th]
	\centering
	\includegraphics[trim=2.4cm 4.2cm 6.3cm 2.5cm, clip=true, width=0.94\linewidth]{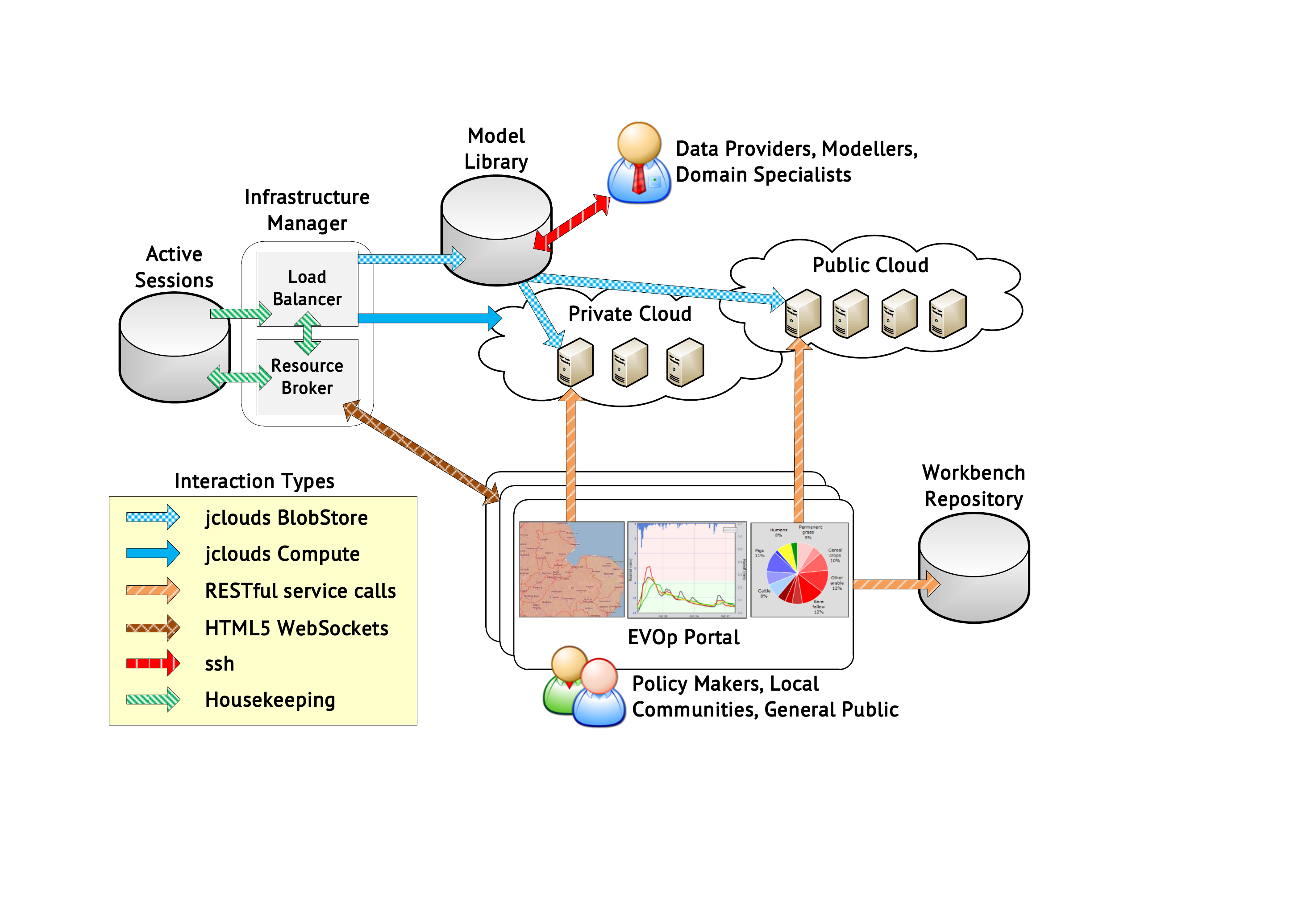}
	\caption{The EVOp Infrastructure}
	\label{fig:interactions}
\end{figure}

Beside providing various information surrounding certain environmental topics, the portal gives users the capability to execute computations on demand in the cloud. This is done using the datasets and models made available through EVOp. \figurename{} \ref{fig:interactions} depicts the EVOp infrastructure and how user interaction is dealt with to enable such transparent computations.

The Model Library is populated by partner domain specialists (e.g. hydrologists, biogeochemists, etc.) in liaison with data providers. The process starts with offline calibration and testing of a model against a certain dataset (e.g. TOPMODEL on the rainfall data of the Eden catchment in the north west of England). The outcome of this process is a virtual machine image optimized to run a fine tuned set of models (exposed as RESTful web services) and equipped with all related data. This streamlined execution bundle is then stored in the Model Library to be instantiated upon demand. If required, an image could be updated to include more historical data or to adjust the implementation of a model in some way.

Once a user navigates to one of the modelling pages within the EVOp portal, a connection is created with the Resource Broker (RB) module of the Infrastructure Manager. RB responds with an address of a cloud instance that is suitable for the type of computation required, along with some session information. Using HTML5 WebSockets for this connection facilitates asynchronous duplex communication without the need for polling or streaming. This reduces network overhead and browser memory usage, and enables RB to manipulate the user session more efficiently. For instance, this is used to update the set of active sessions in order to balance load by sensing when user sessions end. It also allows RB to push any session update to the user browser, such as in the case of migrating the user to a new cloud instance.

The Load Balancer (LB) monitors the health status of running instances with two objectives: minimize costs and maintain instance responsiveness. To minimize cost, user requests are served by default using private cloud instances. Upon saturation of private cloud resources, public cloud instances are used. Reverse migration is undertaken upon detecting underuse of the private cloud. For the latter objective, instance statistics are observed, namely CPU utilization, disk reads and writes, and network usage. Degradation in these metrics, such as sustained high CPU utilization or zero outbound network usage whilst receiving inbound traffic, triggers LB into starting a new instance and redirecting users that were being served by the old instance to the new one. LB also monitors the state of active user sessions and redistributes users on running cloud instances accordingly. RB is used to push updated session information in order to redirect user calls.

The Infrastructure Manager in itself, comprising of the RB and LB modules, is stateless and executed as a cloud instance off a virtual machine image for easy recovery in case of failure. The Active Sessions cache is independent and persistent.

\section{Development \& Deployment}
\label{sec-deploy}

\subsection{Hardware Resources}
\label{sec-deploy-hw}
The infrastructure is deployed as a hybrid consisting of both private resources, managed by Eucalyptus Community Cloud (ECC), and public resources, provided by Amazon Web Services (AWS). ECC provides an open source alternative to the AWS products EC2 (utility computing) and S3 (storage service). This makes it possible, at least in theory, to use the same virtual machine images to start instances in either cloud.

\subsection{Portal Interface}
\label{sec-deploy-portal}
Portal users, including scientists, are not expected to be IT experts and hence would rather not tussle with compatibility issues, security restrictions, etc. Therefore, we developed an intuitive user interface that is tested with stakeholders to ensure a low entry barrier for all targeted user groups.

The use of more than one visualization tool, including graphs and maps, is essential. This is due to the nature of the data, as in the case of rainfall which is a geospatially distributed time series. It also helps in bringing out relationships or patterns within the data, such as correlation between rainfall and runoff levels.

\subsection{Development Cycle}
\label{sec-deploy-agile}
Requirement collection for the EVOp portal was not a trivial task. This is due to the novelty of the application and the need to create highly customized web tools. Furthermore, the EVOp team included researchers from different backgrounds (environmental, computer and social science). Frequent meetings were required to discern any changes that need to be done at an early phase.

\begin{figure}[!th]
	\centering
	\includegraphics[width=\linewidth]{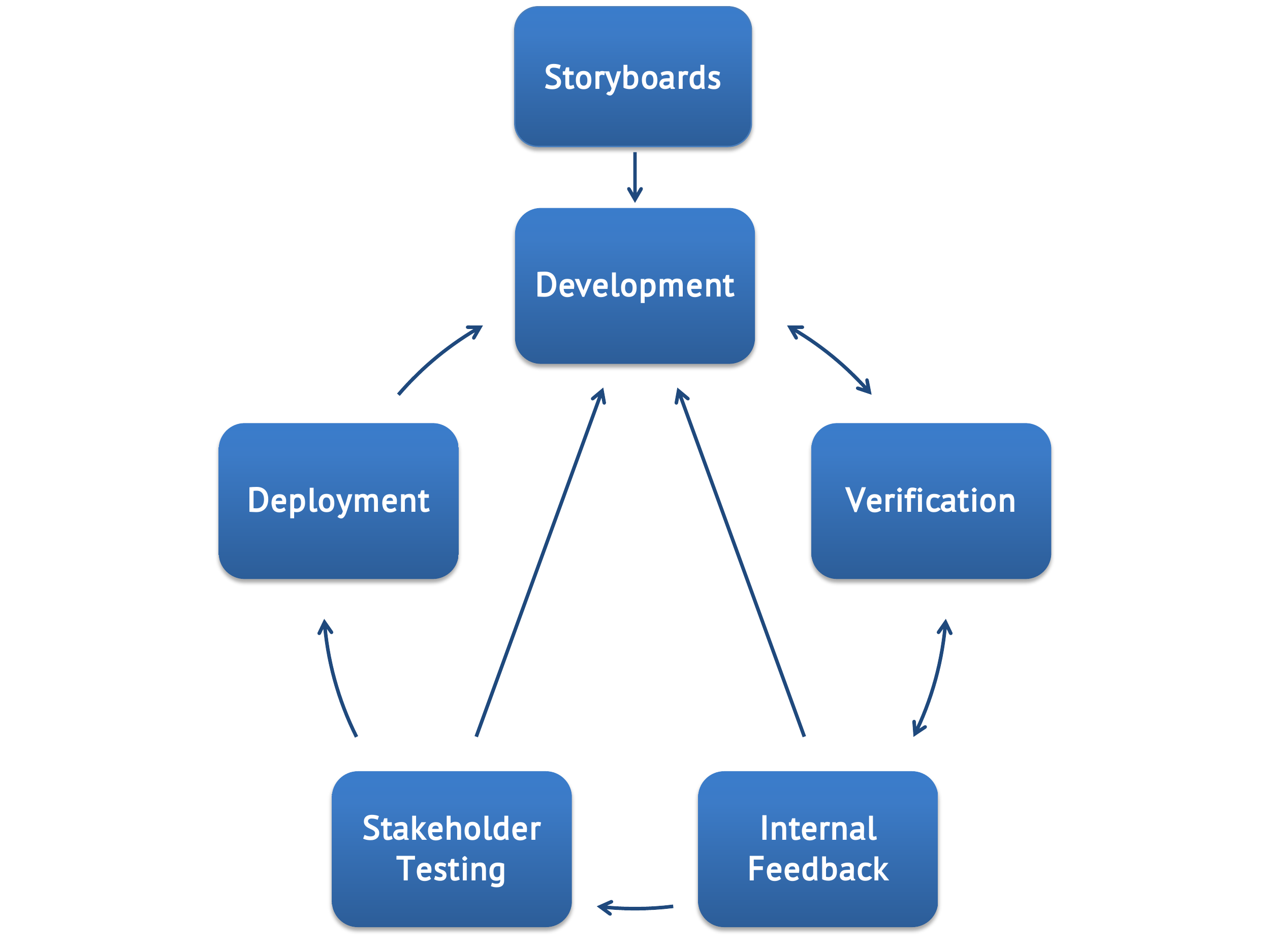}
	\caption{Behaviour-Driven Development Cycle}
	\label{fig:reflections:bdd}
\end{figure}

Therefore, the EVOp development process heavily depended on incremental development and frequent verification. We used the agile-based behavior-driven development methodology (see \figurename{} \ref{fig:reflections:bdd}). \textit{Storyboards}, i.e. a stepped illustration of a fully defined user scenario, are outlined by partner domain specialists (referred to as the \textit{storyboard owners}). Core requirements are drawn from the storyboards. Prototypes are developed based on these requirements, and are iteratively improved and built upon following processes of verification (within the development team, and the storyboard owners) and validation (the wider project consortium, and finally the stakeholders).

\subsection{Prototypes}
\label{sec-deploy-prototypes}
We have thus far worked on developing custom visualization tools for two storyboards. 
\subsubsection{Flooding (at a local scale)}
\begin{figure}[!t]
	\centering
	\includegraphics[width=\linewidth]{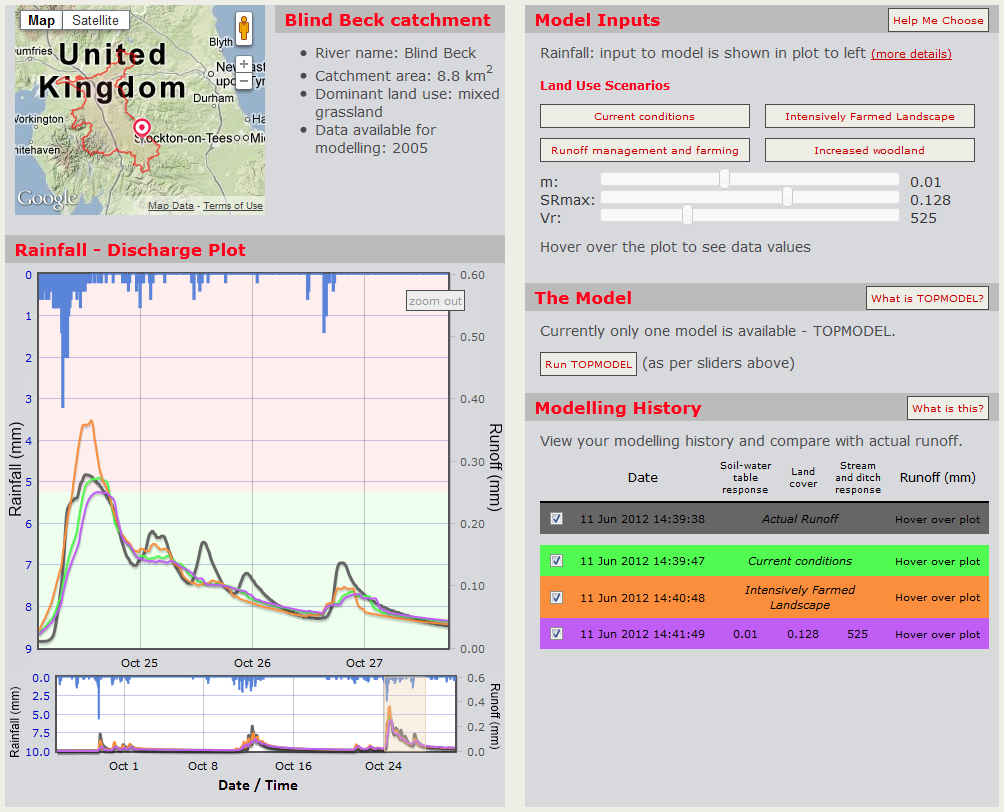}
	\caption{Local Flooding Prototype}
	\label{fig:prototypes:flooding}
\end{figure}

We simulate hydrological interactions to determine where saturated land-surface areas develop, which roughly translates to flood risk areas. Model parameter values (resembling catchment characteristics) could be specified either explicitly by value, or indirectly using one of several predefined land use scenarios. \figurename{} \ref{fig:prototypes:flooding} shows a screenshot from the working prototype for this storyboard.

\subsubsection{Diffuse Pollution (at a national level)}
\begin{figure}[!th]
	\centering
	\includegraphics[width=\linewidth]{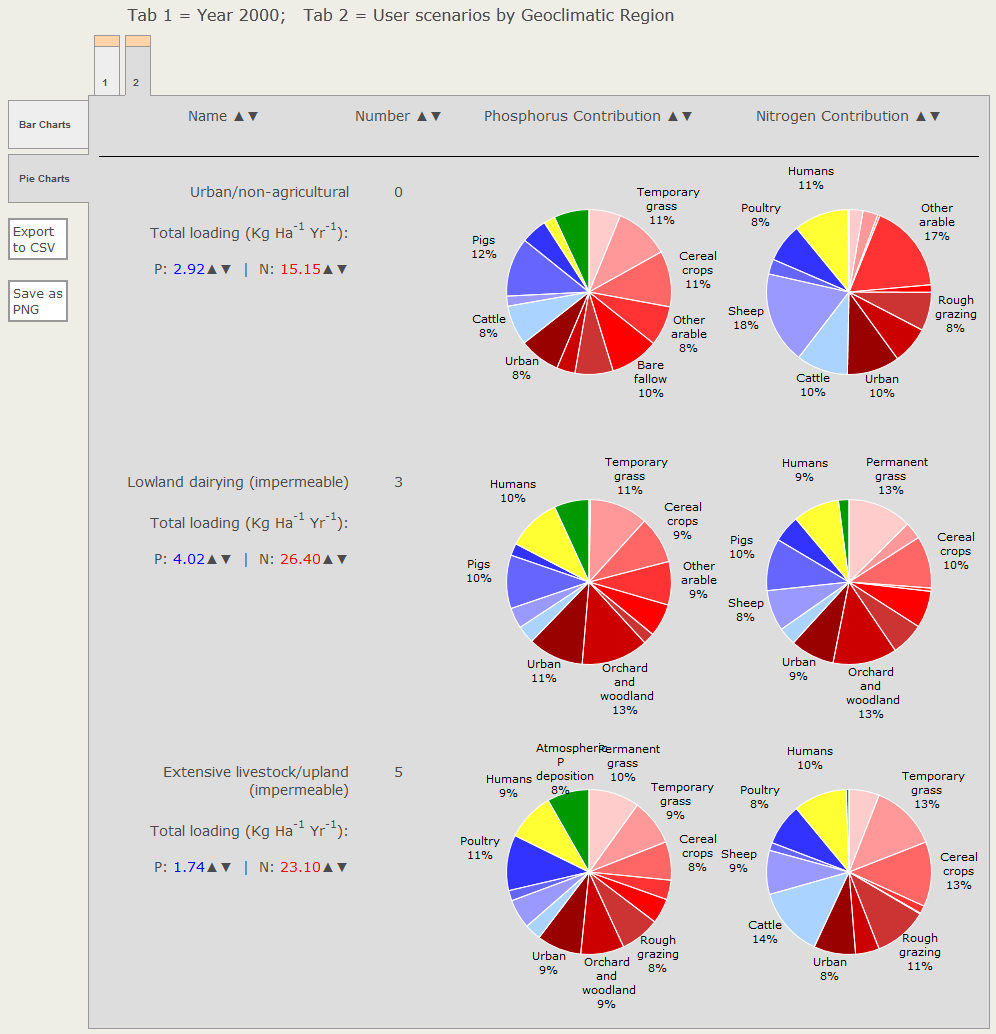}
	\caption{National Diffuse Pollution Prototype}
	\label{fig:prototypes:diffpoll}
\end{figure}

We investigate the diffuse of agricultural pollution through studying the flux of Nitrogen and Phosphorus nutrients from land (at various scales of drainage and reporting units) to rivers and coastal regions. This is particularly useful to explore the impact of existing policy instruments or risk from future environmental changes on the levels of Nitrogen and Phosphorus flux. \figurename{} \ref{fig:prototypes:diffpoll} presents a screenshot from the working prototype for the diffuse pollution storyboard.

\section{Reflections}
\label{sec-reflections}
This section reflects on our experiences while implementing and deploying the EVOp infrastructure. It reviews the lessons learned and difficulties faced, both technical and otherwise.

\subsection{Lessons Learned}
\label{sec-reflections-lessons}
The following are some of the lessons learned, ordered first according to the corresponding design decision.

\subsubsection{Everything as a Service}
\label{sec-reflections-XaaS}
The transparency that comes with XaaS offers a great deal. For \textit{soft} assets, it offers versatile resource management, allowing EVOp to support data assets of different origins: in situ gauging stations, warehoused data stores, and external sources. It also promotes a mashup culture where resources can be shared and reused. We plan to exploit this by facilitating users to connect components (such as models) that were previously living in isolation in order to create more advanced modelling capabilities, as will be discussed in Section \ref{sec-future}.

For tangible resources, XaaS essentially is Infrastructure as a Service (IaaS) where hardware resources could be arranged as and when required. This provisioning of hardware resources as a utility offers elasticity, whereby the infrastructure is allowed to scale to meet user demand and maintain an acceptable quality of service. Consider for instance uncertainty analysis where a model is repeatedly executed using ranges of values for input parameters in order to compensate for any sources of error in how well the data represents the real variables, e.g. topographical representation of a river catchment. This requires substantially more computational resources than a single execution. By providing such resources on demand, IaaS presents such a great advantage when compared to both grid and cluster computing where usage quotas are a common hidnrance for resource-intensive computations.

\subsubsection{RESTful Web Services}
\label{sec-reflections-REST}
SOAP web services require high communication and operation overheads. Graceful service degradation and service migration are complicated due to the need to maintain service state. Performance and scalability also stand to suffer for the same reason.

Adopting RESTful services draws a clear line between the client and the server (i.e. the machine hosting a web service to process a request) which has a huge knock on effect on the scalability and manageability of the infrastructure. As application state is not maintained by the server, there is much less load on it. The client, however, can invoke the server as much as required to change the state throughout the steps of a scientific experiment, the different runs of a simulation, etc. Moreover, this greatly simplifies complicated infrastructure management tasks such as load balancing and failure recovery. In order to optimize performance, end user requests are routed to any available hosted service regardless of current state or previous interactions. Similarly, failed virtual machines are easily replaced by others on another rack, data center, or even cloud provider. Hence, migration does not require advance resource reservation, shared block devices, or any similar techniques.

Consequently, we find the RESTful approach to architect web services very suitable for different types of scientific applications, especially embarrassingly parallel ones such as Monte Carlo simulations, parameter sweeps, uncertainty analysis, etc. where there is no need to share state between transactions.

\subsubsection{Virtualization}
\label{sec-reflections-virt}
Using platform as a service (PaaS) is suitable for applications with fairly consistent execution requirements. However, building custom virtual machines starting from the IaaS level maximizes flexibility, putting very little limitations on the application and removing many of the barriers around development. This engenders an inclusive attitude where scientists with different ways of working within and outside a single discipline can build services using whatever mix of platform and software (data management, processing, statistical, messaging, etc.) they are comfortable with. For example, one scientist might choose a stack comprising of R, Python and PostgreSQL which allows him to run geospatial indexing, while his colleague might prefer to use Hadoop over a NoSQL dataset to parametrically analyze the uncertainty intervals in her experiment. In any case, all they need to provide from an EVOp point of view is a RESTful service that can be invoked through the portal. 
To draw a comparison, scientists using the grid for their computations are tied in to too many specifications: hardware architecture, runtime environment, scheduling interface, and supported application interface. As such, only certain types of jobs can be submitted and precompiling is an unavoidable chore to ensure compatibility.

\subsubsection{Federated Cloud}
\label{sec-reflections-federated}
Using jclouds to provide interoperability across different cloud APIs has cost us slightly more development time. It, however, has been a worthwhile investment. Implementing our infrastructure manager using jclouds saved us from reimplementing large chunks of code when changing the IaaS solution as described below.

Although ECC provides an open source solution that mimics AWS's core services, it requires substantially more operational overhead than we could afford. First, moving a virtual machine image between ECC and AWS is not as easy as one would expect as it needs a lot of preparation before it could be converted and imported. Second, ECC versions 1.6+ suffer from recurring stability issues due to Java memory leaks and other bugs (e.g. server certificate verification failure in versions 2.0+). Unfortunately, the ECC community support was weak, at least when we faced such issues (mid 2011 - early 2012). Indeed, Eucalyptus Systems recently recognized this and moved the community support forum to a new platform called Engage. However, there still remains a huge number of open threads with unresolved issues and unanswered queries, significantly more than what is usually encountered with open source communities.

For these reasons, we decided to switch our private cloud infrastructure manager to OpenStack, also AWS-compatible. OpenStack is still far from being a mature solution and its documentation is rather patchy, but it has a vibrant growing community that provides ample support.

Using a cross-cloud API such as jclouds is also useful when the infrastructure utilization model needs to be adjusted. For example, changing the routing mechanism from `all computations on private cloud till saturation' to something more selective such as `streamlined models to AWS and experimental ones to the private cloud'. Another obvious reason that became clearer through EVOp work is that it is necessary to have a federated open approach as it is impossible to commit the national and international environmental science community to any one provider.

\subsubsection{2-Tier not 3-Tier}
\label{sec-reflections-tiers}
The 2-tier architecture we use for running models has several advantages: less latency between tiers, no performance bottlenecks caused by competition for shared data resources, and less data inconsistency concerns. However, this can only be applied to applications where the data requirements are of small chunks. If the data to be processed becomes too big or does not conform to `write rarely, read frequently', then the logic and data tiers have to be divided. 

\subsubsection{Agile Participatory Development}
\label{sec-reflections-agile}
Defining storyboards and implementing them during relatively short agile development cycles has proven very successful. 
Prototypes are developed and are iteratively improved following processes of verification (within the development team) and validation (with the storyboard owners, the wider consortium, and finally the stakeholders). Such approach is participatory at all stages which made us ``fail early, and fail often'' meaning that changes in plan were frequent, but were low-cost due to the prompt identification of issues with the portal. It also lead to early and frequent engagement with the end users, providing useful feedback and resulting in a set of tools that are relevant and useful to the end users.

\subsection{Difficulties}
\label{sec-reflections-difficulties}
Adopting the RESTful style for web services was not without difficulties. Some scientific communities, such as the geospatial analysis community, specify their domain standards using WS-* SOAP services. 
This meant that in a couple of instances we had to divert from domain standards in order to preserve a RESTful architecture. Service syntax was preserved when veering away from domain standards. Currently, there are discussions in some circles about amending standards to accommodate RESTful style services, e.g. the Open Geospatial Consortium's Web Processing Service (WPS) 2.0 Standards Working Group \cite{wps20swg}.

Furthermore, RESTful service interfaces are by definition simple and uniform across resources. This helps in bringing the aforementioned benefits. However, this also means that any resource-specific actions and associated semantics are lost from the interface to be specified in the service call payload. Thus, the RESTful style does not support inherent self-describing service semantics as SOAP does. Semantic annotations are essential for discovery of and interaction with other web services offering datasets or models. We experimented with several tools of extracting annotations from model or dataset, and used WSDL 2.0 \cite{chinnici2007web} to annotate our services.

We also faced some non-technical difficulties. Of the involved parties, end users are of the first to recognize the potential advantages of moving scientific data and services to the cloud. Previously, a scientist typically needed to have the data on her computer, a step of often underestimated difficulty. She then has to find or develop a model, and proceed to calibrate and run it. The model results are examined with a possibility of repeating most of this cycle over and over again. Therefore, such users immediately identify the ease of use, universal access, and abundance of resources that comes with a cloud infrastructure.

However, other stakeholders have different perceptions. For instance, some data producers are apprehensive about providing their data assets through what they perceive as new, untested means. This is a tough problem and can only be resolved, if at all, by educating data owners about cloud computing. This would provide some assurance about the flow of data through the infrastructure and explain that public cloud providers, the target of most security apprehensions, have whole expert teams working on security to honor their SLAs. Success stories, such as EVOp and Eduserv (a non-profit organization that provides technology solutions, including IaaS provisions, to the education, health and public sectors \cite{eduserv}), could also help alter attitudes.

Other non-technical difficulties include payment systems and organizational politics. \cite{McGough12experiences} highlights some of these.

\section{Future Work}
\label{sec-future}
We are active in pursuing additional data sources, both historical and real time, and accordingly develop suitable curation processes and visualization tools. We are also planning to expand the spectrum of tools offered by EVOp by supporting more domain specialists to popoulate the Model Library with more images. More importantly, we are looking to increase the room for customization by supporting workflow execution. So far, we have been building web based prototypes based on very specific use cases outlined by storyboards. A workflow is a conglomerate scientific process composed of a directed acyclic graph of basic execution units (e.g. executables, scripts, web services, etc.). Workflows allow `advanced' users (i.e. domain specialists from the scientific or govermental communities) to create complex experiments that can be easily tweaked and replayed. This offers reproducibility and traceability. If described in a standard way, a workflow can be shared and reused by others in order to build upon it, reproduce results, or compare techniques. Indeed, sharing workflows has proven to be quite useful in other fields of science such as bioinformatics. Observing established work in this area, such as Taverna \cite{Oinn2006taverna}, will allow us to leverage associated platforms, such as myExperiment \cite{goble2010myexperiment}, to disseminate workflows and create collaborative communities.

To this end, we are working on providing workflow composition through a web browser using the lightweight graphical user interface presented in \cite{KISS}. We have also added to the Model Library a generic virtual machine image with a workflow execution manager. For such web service orchestration, constituent web services need to uphold behavioral contracts othrwise interacting with different loosely-coupled services would incur a high level of uncertainty. The two important contracts that concern us here relate to formalizing interactions and supporting monitoring (either polling, push-based, or both). Graceful service degradation is, however, not important at the atomic constituent RESTful service level as substitution and relocation is simple enough and sufficient. Nonetheless, assurance of graceful degradation is needed if orchestration includes external services not directly controlled by EVOp.

Beyond the lifetime of the pilot phase of the project, we are looking to expand beyond water related science into other areas such as soil science and biodiversity.

\section{Related Work}
\label{sec-related}

Scientific research has been aided by cyberinfrastructures for quite some time. Different distributed paradigms, such as HPC and grid computing, have been used over the years to build such infrastructures. Currently, there are a growing number of efforts to enable scientific research using cloud infrastructuress. These range from generic research support web tools to domain-specific virtual research environments. Examples of the former include testing IaaS solutions (e.g. EmuLab \cite{emulab}), statistical analysis platforms (e.g. Biocep-R \cite{Chine2009BiocepR} and CloudNumbers \cite{cloudnumbers}), and social networks (e.g. Mendeley and Academia.edu).

Efforts for designing domain-driven solutions include the following two architectural proposals. \cite{schaffer10towards} presents a use case of similar architectural elements and hybrid infrastructure deployment, but does not use RESTful services. \cite{Roth11vre} defines a generic high-level framework for assembling virtual research environments.

Domain-driven solutions are used for data discovery, data normalisation, and workflow execution. In the domain of environmental and geosciences, the NSF-funded Consortium of Universities for the Advancement of Hydrologic Science (CUAHSI) developed several tools to enable access to water-related research and data. Their Hydrologic Information System (HIS) provides unified access to data, tools and models relating to hydrological research. The HIS index can be accessed by water-related federated search engines and dataset repositories such as NWIS \cite{goodall2008first} and STORET \cite{STORET}. 
Additionally, the Community Hydrologic Modeling Platform (CHyMP) \cite{Famiglietti11} allows the development, support and sharing of models to serve the hydrologic community using pre-existing modelling technologies such CSDMS and NASA LIS.

Other efforts include the Penn State Integrated Hydrologic Model (PIHM) \cite{leonard2010data} which presented a prototype of orchestrating terrestrial watershed models in order to predict water distribution. PIHM is envisioned to move to a cloud infrastructure in the very near future. More recently, EarthCube \cite{jacobs2012earthcube} has emrged as a collaborative effort to create a cloud-based virtual research environment to share data and knowledge about geosciences. In the commercial sector, ESRI is offering a number of cloud-based geospatial services, such as ArcGIS Online and AWS-ready ArcGIS Server.

\section{Conclusion}
\label{sec-conclusion}
In this paper we presented EVOp, a cloud-based virtual observatory for environmental science. EVOp provides web access to data and tools that help different stakeholders in engaging with pressing environmental issues. The underlying infrastructure is a tailored hybrid cloud consisting of owned and leased hardware resources. The essence of the EVOp architecture is to focus on assets rather than on transactions. From this stems the importance of representing all resources through a uniform interface.

We presented the lessons learned through our experience. We learned that representing all resources as a service enables easy management and provides opportunities for integration, encouraging a mashup culture. We also learned that adopting the RESTful architectural style reduces management overheads. RESTful services are stateless which makes them easy to create and recreate (for fault mitigation). Lack of state also makes them easy to operate and scale. However, they do not offer much in terms of self-description through their interface. We also experimented with federated clouds, which offers more options in terms of management and cost control, and prevents vendor lock in. Finally, we faced some non-technical difficulties, such as perceptions of trust and security surrounding cloud computing in some communities (especially concerning data licensing and IPR). We put forward some measures that could be taken to alleviate such concerns.

Cloud computing offers a lot of potential for science by enabling virtual research environments such as EVOp. A cloud infrastructure offers the flexibility to integrate and support varied resources, and a low entry barrier when compared to previous distributed systems. We therefore expect to see a growth in the number of efforts similar to EVOp in the near future. We hope that the experiences presented in this paper would be useful to such efforts.

\bibliographystyle{abbrvnat}
\bibliography{CloudDP}

\begin{thebibliography}{21}
\providecommand{\natexlab}[1]{#1}
\providecommand{\url}[1]{\texttt{#1}}
\expandafter\ifx\csname urlstyle\endcsname\relax
  \providecommand{\doi}[1]{doi: #1}\else
  \providecommand{\doi}{doi: \begingroup \urlstyle{rm}\Url}\fi

\bibitem[KIS()]{KISS}
{The KISS workflow designer}.
\newblock \url{http://dev.mygrid.org.uk/wiki/display/tav/KISS}.

\bibitem[clo()]{cloudnumbers}
{cloudnumbers.com - High Performance Computing (HPC) in the Cloud}.
\newblock \url{http://cloudnumbers.com/}.

\bibitem[edu()]{eduserv}
{Eduserv - Public Sector IT Specialist}.
\newblock \url{http://www.eduserv.org.uk/cloud}.

\bibitem[emu()]{emulab}
{Emulab - Network Emulation Testbed Home}.
\newblock \url{http://emulab.net/}.

\bibitem[jcl()]{jclouds}
{jclouds - multi-cloud library}.
\newblock \url{http://code.google.com/p/jclouds/}.

\bibitem[wps()]{wps20swg}
{Web Processing Service 2.0 SWG}.
\newblock \url{http://www.opengeospatial.org/projects/groups/wps2.0swg}.

\bibitem[Blair and El-khatib(2011)]{blair2011cloud}
G.~S. Blair and Y.~El-khatib.
\newblock {A Cloud-based Virtual Observatory for Environmental Science}.
\newblock \emph{OpenWater Symposium}, page 102, April 2011.

\bibitem[Chine(2009)]{Chine2009BiocepR}
K.~Chine.
\newblock {Scientific Computing Environments in the Age of Virtualization
  Toward a Universal Platform for the Cloud}.
\newblock In \emph{IEEE International Workshop on Open-source Software for
  Scientific Computation (OSSC)}, pages 44--48, September 2009.

\bibitem[Chinnici et~al.(2007)Chinnici, Moreau, Ryman, and
  Weerawarana]{chinnici2007web}
R.~Chinnici, J.-J. Moreau, A.~Ryman, and S.~Weerawarana.
\newblock \emph{{Web Services Description Language (WSDL) version 2.0 part 1:
  Core Language}}.
\newblock W3C Recommendation, 20070626 edition, June 2007.

\bibitem[Famiglietti et~al.(2011)Famiglietti, Murdoch, Lakshmi, Arrigo, and
  Hooper]{Famiglietti11}
J.~S. Famiglietti, L.~Murdoch, V.~Lakshmi, J.~Arrigo, and R.~Hooper.
\newblock {Establishing a Framework for Community Modeling in Hydrologic
  Science}.
\newblock \url{http://www.cuahsi.org/chymp.html}, March 2011.

\bibitem[Fielding(2000)]{fielding2000architectural}
R.~T. Fielding.
\newblock \emph{{Architectural Styles and the Design of Network-based Software
  Architectures}}.
\newblock PhD thesis, University of California, Irvine, 2000.

\bibitem[Goble et~al.(2010)Goble, Bhagat, Aleksejevs, Cruickshank, Michaelides,
  Newman, Borkum, Bechhofer, Roos, Li, and De~Roure]{goble2010myexperiment}
C.~A. Goble, J.~Bhagat, S.~Aleksejevs, D.~Cruickshank, D.~Michaelides,
  D.~Newman, M.~Borkum, S.~Bechhofer, M.~Roos, P.~Li, and D.~De~Roure.
\newblock {myExperiment: A Repository and Social Network for the Sharing of
  Bioinformatics Workflows}.
\newblock \emph{Nucleic Acids Research}, 38\penalty0 (suppl 2):\penalty0
  W677--W682, July 2010.

\bibitem[Goodall et~al.(2008)Goodall, Horsburgh, Whiteaker, Maidment, and
  Zaslavsky]{goodall2008first}
J.~L. Goodall, J.~S. Horsburgh, T.~L. Whiteaker, D.~R. Maidment, and
  I.~Zaslavsky.
\newblock {A First Approach to Web Services for the National Water Information
  System}.
\newblock \emph{Environmental Modelling \& Software}, 23\penalty0 (4):\penalty0
  404--411, 2008.

\bibitem[Jacobs(2012)]{jacobs2012earthcube}
C.~Jacobs.
\newblock {A Vision for, and Progress Towards EarthCube}.
\newblock In \emph{European Geosciences Union General Assembly Conference
  Abstracts}, volume~14, page 1227, April 2012.

\bibitem[Leonard et~al.(2010)Leonard, Duffy, and Bhatt]{leonard2010data}
L.~N. Leonard, C.~Duffy, and G.~Bhatt.
\newblock {Data-intensive Hydrologic Modeling: A Cloud strategy for integrating
  PIHM, GIS, and Web-Services}.
\newblock In \emph{American Geophysical Union Fall Meeting Abstracts},
  volume~1, page~8, December 2010.

\bibitem[Malkowski et~al.(2009)Malkowski, Hedwig, and Pu]{Malkowski09workload}
S.~Malkowski, M.~Hedwig, and C.~Pu.
\newblock {Experimental Evaluation of N-tier Systems: Observation and analysis
  of multi-bottlenecks}.
\newblock In \emph{IEEE International Symposium on Workload Characterization
  (IISWC)}, pages 118--127, October 2009.

\bibitem[McGough et~al.(2012)McGough, Glenis, Kilsby, Kutija, and
  Wodman]{McGough12experiences}
S.~McGough, V.~Glenis, C.~Kilsby, V.~Kutija, and S.~Wodman.
\newblock {Experiences in running High Throughput Computing on the Cloud}.
\newblock In \emph{OGF Workshop on Science Applications and Infrastructure in
  Clouds and Grids}, March 2012.

\bibitem[Oinn et~al.(2006)Oinn, Greenwood, Addis, Alpdemir, Ferris, Glover,
  Goble, Goderis, Hull, Marvin, Li, Lord, Pocock, Senger, Stevens, Wipat, and
  Wroe]{Oinn2006taverna}
T.~Oinn, M.~Greenwood, M.~Addis, M.~N. Alpdemir, J.~Ferris, K.~Glover,
  C.~Goble, A.~Goderis, D.~Hull, D.~Marvin, P.~Li, P.~Lord, M.~R. Pocock,
  M.~Senger, R.~Stevens, A.~Wipat, and C.~Wroe.
\newblock {Taverna: Lessons in creating a workflow environment for the life
  sciences}.
\newblock \emph{Concurrency and Computation: Practice \& Experience},
  18\penalty0 (10):\penalty0 1067--1100, August 2006.
\newblock ISSN 1532-0626.

\bibitem[Roth et~al.(2011)Roth, Hecht, Volz, and Jablonski]{Roth11vre}
B.~Roth, R.~Hecht, B.~Volz, and S.~Jablonski.
\newblock {Towards a Generic Cloud-Based Virtual Research Environment}.
\newblock In \emph{IEEE 35th Annual Computer Software and Applications
  Conference Workshops (COMPSACW)}, pages 267--272, July 2011.

\bibitem[Sch{\"a}ffer et~al.(2010)Sch{\"a}ffer, Baranski, and
  Foerster]{schaffer10towards}
B.~Sch{\"a}ffer, B.~Baranski, and T.~Foerster.
\newblock {Towards Spatial Data Infrastructures in the Clouds}.
\newblock \emph{Geospatial Thinking}, pages 399--418, 2010.

\bibitem[{US Environmental Protection Agency, Office of Water}()]{STORET}
{US Environmental Protection Agency, Office of Water}.
\newblock {STOrage and RETrieval: The US EPA Water Quality Database}.
\newblock \url{http://www.epa.gov/storet/}.

\end{thebibliography}

\end{document}